# Telecommunication-wavelength two-dimensional photonic crystal cavities in a thin single-crystal diamond membrane

Kazuhiro Kuruma[1,a], Afaq Habib Piracha[1], Dylan Renaud[1], Cleaven Chia[1], Neil Sinclair[1,2], Athavan Nadarajah[3], Alastair Stacey[3,4], Steven Prawer[3], and Marko Lončar[1]

[1]*John A. Paulson School of Engineering and Applied Sciences, Harvard University, Cambridge, Massachusetts 02138, USA*

[2]*Division of Physics, Mathematics and Astronomy, and Alliance for Quantum Technologies (AQT), California Institute of Technology, Pasadena, California 91125, USA*

[3]*School of Physics, University of Melbourne, Victoria 3010, Australia*

[4]*School of Science, RMIT University, Melbourne, Victoria, 3001, Australia*

[a] E-mail: kkuruma@seas.harvard.edu

**Abstract**

**We demonstrate two-dimensional photonic crystal cavities operating at telecommunication wavelengths in a single-crystal diamond membrane. We use a high-optical-quality and thin (~ 300 nm) diamond membrane, supported by a polycrystalline diamond frame, to realize fully suspended two-dimensional photonic crystal cavities with a high theoretical quality factor of ~ $8\times10^6$ and a relatively small mode volume of ~2 $(\lambda/n)^3$. The cavities are fabricated in the membrane using electron-beam lithography and vertical dry etching. We observe cavity resonances over a wide wavelength range spanning the telecommunication O- and S-bands (1360 nm-1470 nm) with *Q* factors of up to ~1800. Our method paves the way for on-chip diamond nanophotonic applications in the telecommunication-wavelength range.**

Photonic crystal (PhC) cavities with high quality ($Q$) factors and small mode volumes ($V$s), have attracted much attention because of their ability to strongly confine light in time and space. The enhanced light-matter interactions offered by PhC cavities can be utilized for the study of cavity quantum electrodynamics [1], as well as for various photonic applications including on-chip optical interconnects [2], nonlinear optics [3], optomechanics [4], quantum information processing [5], and sensing [6], and atomic physics [7]. Silicon (Si) is one of the leading materials to realize high-$Q$ PhC cavities [8,9] due to its high refractive index, mature fabrication process, and availability of wafers on insulator. However, Si-based PhC devices can suffer from two-photon absorption and related free carrier absorption in the near-infrared region due to the relatively small ~1.1 eV bandgap of Si [10]. To overcome this issue, various wide gap materials such as diamond [11], SiC [12], AlN [13], and $Si_3N_4$ [14] have been investigated. Among them, single-crystal diamond stands out because of its attractive optical and physical properties: relatively large refractive index (2.4), wide transparency window from the ultraviolet to far-infrared, high thermal conductivity (~2200 W ∕ m · K) and small thermo-optic coefficient (~$10^{-5}$ $K^{-1}$). So far, there have been many reports on PhC cavities fabricated in single-crystal diamond. Most of them are focused on operation at visible wavelengths to enhance the interaction between light and diamond color centers such as nitrogen-vacancy (NV) [15–18], silicon-vacancy (SiV) [19–22] and tin-vacancy (SnV) centers [23,24]. Recently, diamond PhC cavities operating at telecommunication wavelengths have been demonstrated and utilized for applications in optomechanics [25,26], but these works utilize one-dimensional (1D) PhC cavities.

While diamond 1D PhC cavities have been the workhorse of quantum-photonics for years, 2D PhC cavities may offer additional advantages. For example, the highest $Q$ PhC cavities reported to date are realized using 2D PhCs ($Q$~ $10^7$) [9], more than an order of magnitude larger than in the case of 1D PhCs ($Q$ ~ $10^6$) [8]. For cryogenic applications, 2D PhC cavities can offer better thermal conductivity and heat dissipation, which is one of the challenges that 1D structures are currently facing, particularly when operating at mK temperatures [27,28]. For optomechanics, 2D mechanical crystals used as phononic shields allow increased mechanical cavity $Q$s [29]. Furthermore, 2D structures enable easier integration with other on-chip functionalities, including acoustic control of color centers [30]. Finally, for nonlinear optics applications, the 2D approach also allows more flexibility in designing cavity resonances for, e.g., the realization of ultra-low threshold Raman lasers [31], leveraging diamond's strong Raman nonlinearity [32].

The lack of the telecom-wavelength 2D PhC cavities in diamond could be due to difficulties associated with the fabrication of high-quality and large-area, suspended, diamond planar structures. For example, angle etching [11], one of the leading fabrication approaches, is not compatible with the fabrication of 2D PhCs due to the resulting characteristic triangular cross-sectional profile. Moreover, approaches based on thinning diamond plates (initial thickness of 5-30 $\mu$m) to sub-micron-thick films by reactive ion etching (RIE) [17,18,33] often suffer from large thickness variations (so called "wedging") due to non-uniform thickness of the starting, mechanically-polished material [17], thereby hindering large-scale fabrication of PhC cavities in diamond. Although ion slicing can produce large-area and uniform thin films [34–36], ion damage in these films and residual built-in strain induced by ion implantation are unavoidable [37]. Recently, quasi-isotropic etching [38,39] has been also used to realize 2D PhC cavities in bulk diamond [40]. However, this method requires long etch times to completely undercut larger 2D slab structures, and can introduce backside roughness and the thickness variations of the slab.

In this letter, we fabricate and characterize telecommunication-wavelength 2D PhC cavities in a single-crystal diamond membrane (SCDM). We employ an ultra-thin and homogenous diamond membrane fabricated by a combination of ion implantation, chemical vapour deposition (CVD) overgrowth, and RIE [41]. With the SCDM, we develop a process, which only relies on standard electron beam lithography and dry etching without additional undercut etching, to fabricate a fully suspended 2D PhC cavity with a theoretical $Q$ factor $\sim 8\times 10^6$ and $V \sim 2$ $(\lambda/n)^3$. We measure fabricated PhC cavities by a resonant scattering method, finding cavity resonances with $Q$ factors up to ~1800 in the telecommunication O- to S-bands. Our results show the potential of developing diverse diamond nanophotonic devices based on PhC cavities, especially at telecommunication wavelengths, for various applications, including nonlinear optics and optomechanics.

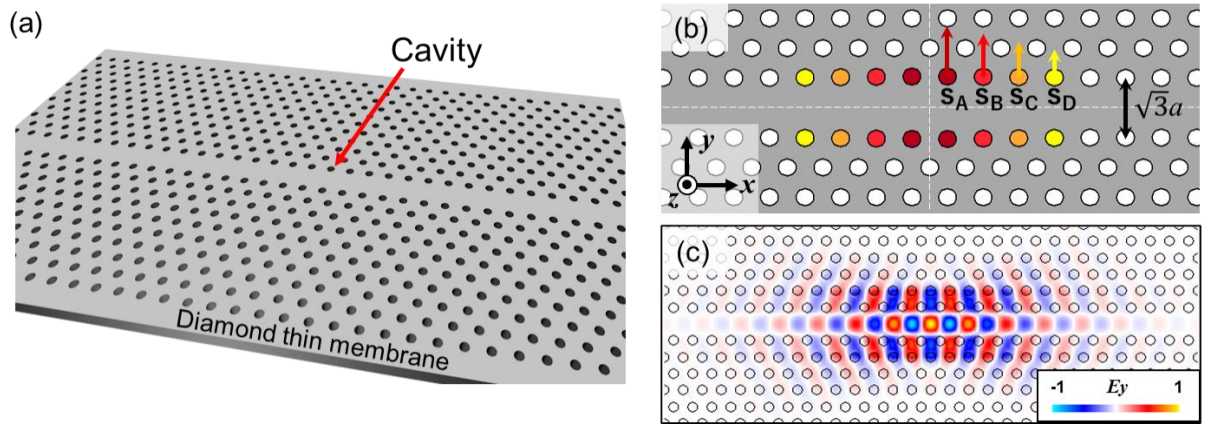

**Fig. 1.** (a) Schematic of our suspended 2D PhC cavity in a thin diamond membrane. (b) Detailed depiction of the air hole shifts. The arrows indicate the shifts of the colored air holes in the $y$ direction, with the magnitude of each shift color-coded (largest in dark red, smallest in bright yellow). The shifts are mirror-symmetrical with respect to the plane across the cavity center (white dashed lines). (c) Calculated electric field ($E_y$) profile of the fundamental cavity mode with a theoretical $Q$ factor of $\sim 8\times 10^6$.

We investigate a width-modulated line-defect PhC cavity [42], as schematically shown in Fig 1(a). The PhC consists of a triangular lattice of air holes in a 300 nm-thick diamond membrane (slab) with a line-defect waveguide formed by a missing row of air holes along the Γ–K direction. The lattice constant ($a$) ranges from 502 to 542 nm, and the air hole radius ($r$) is 135 nm. We introduce shifts of the air holes ($S_A$ = 22 nm, $S_B$ = 16.5 nm, $S_C$ = 11 nm, $S_D$ = 5.5 nm) into the PhC waveguide to form a cavity, as shown in Fig. 1(b). The tapered shifts of the air hole position along the $x$-direction enable the realization of high $Q$ factors even when the shifts are only applied to the air holes in the first row adjacent to the PhC waveguide [42,43]. This cavity design is also advantageous as it allows high $Q$ factors and it is relatively insensitive to structural disorders such as variations of air hole positions and radii, compared to other conventional cavity designs such as L3, and H0 type cavities, except heterostructure cavities [44]. For $a$ = 542 nm and the slab thickness of 300 nm, we obtain a high theoretical $Q$ factor of ~$8\times10^6$ and a moderate $V$ of ~2.15 $(\lambda/n)^3$ for the fundamental cavity mode (wavelength $\lambda$ = 1.52 µm, refractive index of the diamond slab $n$ = 2.4). Note that an increase in the slab thickness to 400 nm results in a theoretical $Q$ factor of ~$1.2\times10^7$ without a large change (< 1%) in $V$ ($\lambda$ = 1.6 µm). The obtained $Q$ and $V$ are comparable to values previously reported for 2D PhC cavities in diamond [45]. Figure 1(c) shows the calculated electric field ($E_y$) distribution of the fundamental cavity mode using a 3D finite-difference time domain (FDTD) method.

The cavity fabrication process is illustrated in Fig 2. A SCDM with a thickness of ~300 nm is prepared using a scalable fabrication technique previously reported in [41,46]. This technique involves a combination of He$^+$ ion implantation and microwave plasma CVD overgrowth to form a high-optical-quality, free-standing, thin membrane from a single-crystal diamond substrate [34,41,47]. We use a thick polycrystalline diamond frame to support the fabricated membrane to enable its handling. The frame is firmly bonded to the SCDM using diamond CVD overgrowth at the edges so that there is no stress due to differential thermal expansion during processing. The removal of the ion-damaged layer and thinning of the SCDM down to the desired thickness are made possible by oxygen-plasma RIE during the SCDM fabrication. The RIE process keeps a low root mean square surface roughness of ~ a few nm [41]. The typical thickness variation of the SCDM is ~100 nm/mm, which is mainly due to imperfect surface polishing after diamond CVD overgrowth. We consider that the remaining thickness variation can be further improved by optimization of the polishing process and technique [48]. We note that color centers such as NV and SiV centers can be created in the SCDM via CVD growth [41] or ion implantation [49].

In this study, we employ a high-pressure high-temperature Type 1b single-crystal diamond substrate as a seed for the CVD overgrowth step. Our SCDM is supported by a 300 µm-thick polycrystalline diamond frame (1.2 mm × 1.2 mm) with nine circular windows (each has a diameter of 240 µm) [Fig. 2(a)]. The pitch of the windows is 350 µm. It is noteworthy that we can easily customize the size of the diamond frame and the configuration of windows (number, size, and shape) akin to that shown in [41]. The SCDM can be prepared with a dimension of 4 mm × 4 mm and the window size can be scaled up to ~ 1 mm × 1 mm without breakage of the membrane.

The inset of Fig. 2(a) shows an optical microscope image of one of the SCDM windows before cavity fabrication. For safer handling of the SCDM and frame during the fabrication process, we first bond the frame to a Si substrate using hydrogen silsesquioxane (HSQ) [Fig. 2 (b)]. It is noted that it is not necessary to use the Si carrier substrate for handling the SCDM. The use of a larger SCDM and frame can also allow easier handling. Before spin-coating an electron beam (EB) resist (ZEP 520A), a 100 nm-thick SiN layer is deposited on the SCDM by plasma-enhanced chemical vapor deposition (PECVD) [Fig. 2 (c)]. We pattern arrays of the designed cavity into the resist by EB lithography followed by resist development using o-xylene [Fig.2 (d)]. The SiN layer serves as a hard mask and is etched by inductively coupled plasma-RIE (ICP-RIE) using sulfur hexafluoride ($SF_6$) and octafluorocyclobutane ($C_4F_8$) gases [Fig. 2 (e)]. After removing the EB resist, we use an oxygen-based ICP-RIE (ICP = 700 W, RF = 100 W, P = 1.3 Pa, and T = 25°C) to fabricate PhC cavities into the SCDM [Fig. 2 (f)]. Finally, we dip the SCDM and frame into hydrogen fluoride (HF) to remove the SiN and HSQ layers. The final SCDM windows with suspended PhC cavities [Fig. 2 (g)] are completed after the frame is detached from the Si substrate by HF. For cleaning the sample surface, we also immersed the SCDM and frame into a piranha solution (3:1 mixture of sulfuric acid and hydrogen peroxide) just after removing the SiN and $Al_2O_3$ layers.

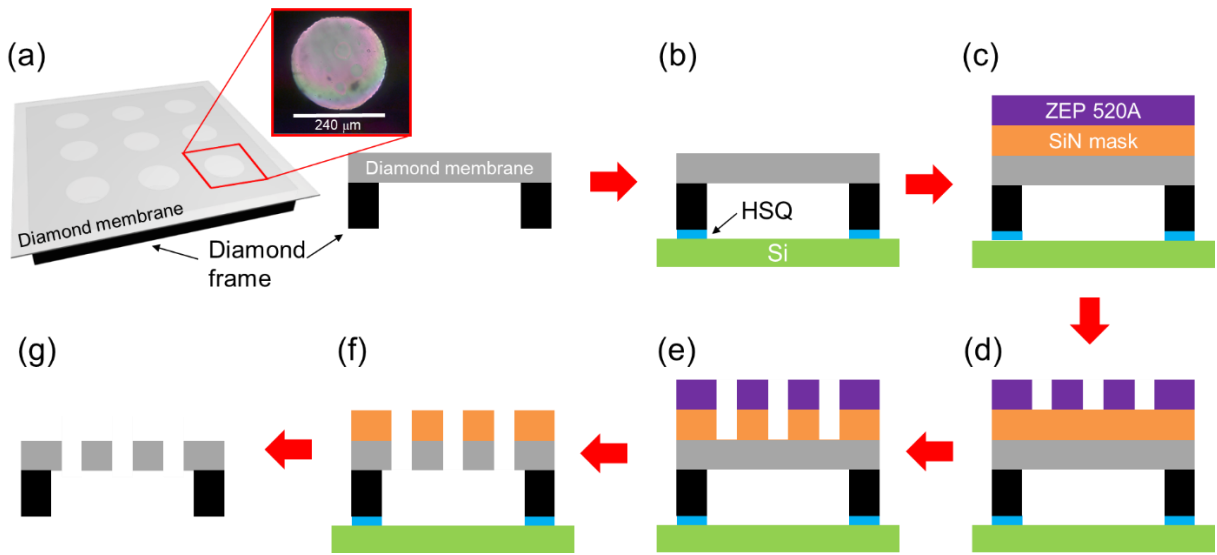

**Fig. 2.** Process flow of the cavity fabrication using a thin diamond membrane. (a) 3D view (left) and the cross-section (right) of the thin membrane on a polycrystalline diamond frame. The inset shows an optical microscope image of one of the thin membrane windows before cavity fabrication. (b) The membrane on the diamond frame is attached to a Si substrate using HSQ. (c) A SiN layer is formed on the membrane by PECVD before spin coating of the EB resist (ZEP 520A). The cavity design is patterned into the thin membrane by (d) EB lithography, (e) RIE of the SiN layer and (f) oxygen plasma RIE using the SiN mask. (f) The diamond membrane is dipped into HF to remove the SiN and HSQ layers. (g) Final structure with suspended 2D PhC cavities.

Figure 3 (a) shows an optical microscope image of one of the SCDM windows with a dense array of 2D PhC cavities. Approximately 50 devices (typical device area is ~350 $\mu m^2$) are fabricated in one SCDM window. We fabricate ~250 devices in total into 6 windows on one SCDM. It is possible to further scale up the number of the devices using larger SCDMs and windows. We do not see any significant breakage of the SCDM window after the cavity fabrication, suggesting our method allows fabrication of diamond nanophotonic devices across the whole SCDM area. The color patterns on the SCDM window are likely due to bending of the membrane in conjunction with thickness variations inside the window. The inset of Fig. 3(a) shows an enlarged view of a PhC cavity and a rectangular-shaped trench which is utilized to measure the slab thickness and confirm that fabricated PhC cavities are still suspended after fabrication. Figures 3(b) and (c) show scanning electron microscopy (SEM) images of an array of the fabricated 2D PhC cavities and a zoom-in view of the cavity, respectively. The thickness variation is measured to be ~ 25 nm over a 240-$\mu$m diameter window using SEM. Among ~250 devices fabricated in 6 windows, approximately half of the devices can be used for optical measurements since the thickness of 3 windows are thinner than the targeted thickness of ~300 nm due to a polishing-induced thickness variation.

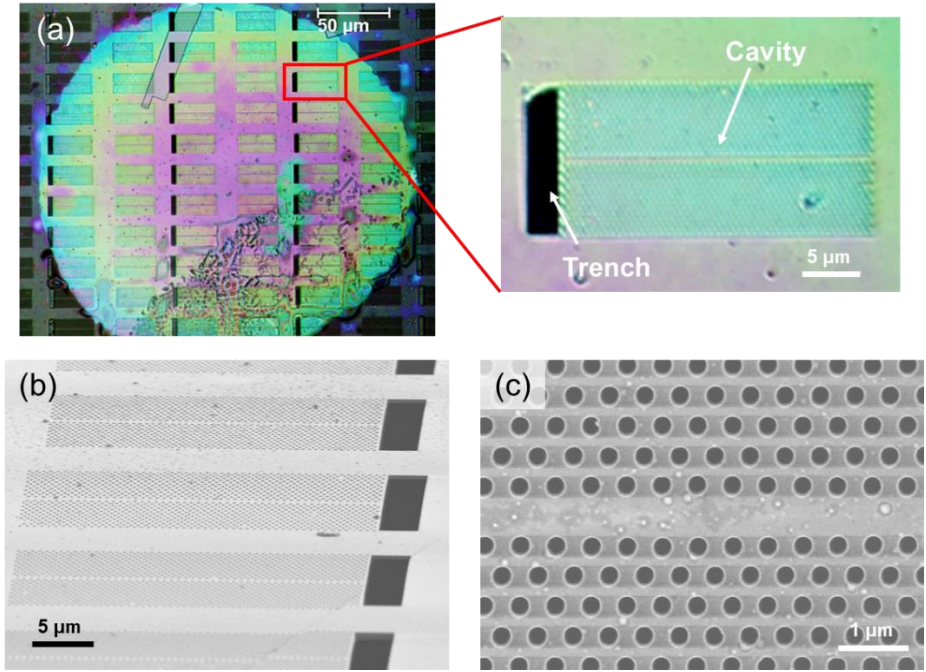

**Fig. 3.** (a) Optical microscope image of an array of fabricated PhC cavities on a thin diamond membrane window. Inset shows an enlarged microscope image of one of the PhC cavities. Scanning electron microscope images showing (b) angled and (c) top-down views of fabricated cavities. The image in (b) is taken at 75° stage tilt.

Next, we characterize the fabricated cavities in the SCDM window shown in Fig. 3 (a) at room temperature using a resonant scattering method [8,50]. The simplified optical setup is depicted in Fig. 4 (a). We use a supercontinuum laser with a spectrum that spans from 410 nm to 2400 nm. The optical output power of the laser is fixed to ~20 mW. The laser light is focused onto the sample by an objective lens (OL) with a 100× magnification and a 0.5 numerical aperture. The reflected light from the sample is collected via the same OL and sent to an optical spectrum analyzer (OSA, resolution ~0.2 nm). In order to extract the cavity signals from the reflected light, we use two linear polarizers that are set to be in front and behind a beam splitter. After fixing the polarization of the laser light by the front polarizer, the reflected light is polarized orthogonal to the input light by the other polarizer, which eliminates the backscattered laser light. Figure 4(b) shows the reflectance spectrum of a fabricated cavity with lattice constant $a = 542$ nm. The sharp resonance at 1470.1 nm likely originates from the PhC cavity mode spatially localized at the cavity center, which is confirmed by varying the location of laser excitation. To further investigate the observed cavity mode, we measure the $a$-dependence of the resonance wavelength. Figure 4 (c) shows the summary of the measured peak wavelength for PhC cavities with $a$ varying between 502 and 542 nm. We observe cavity resonances over a broad range of wavelengths from 1360 to 1470 nm. As expected from simulation (shown later), the resonance wavelength increases as $a$ becomes larger, which is consistent with observations from other PhC cavities [51,52]. Moreover, the experimental data is well-matched to the calculated resonance wavelengths of the fundamental cavity mode. The calculation is accomplished by FDTD simulations considering the measured air hole radius and slab thickness by SEM. These results suggest that the observed resonance originates from the cavity mode shown in Fig 1(c). In addition, we confirmed that PhC cavities with similar lattice constants, fabricated in a different window, also exhibit cavity resonances at telecommunication wavelengths. To evaluate $Q$ factors for our fabricated cavities, we fit the observed spectra with a Lorentzian function, as shown in Fig. 4(b). While typical $Q$ factors of our fabricated cavities are less than $10^3$, the cavity shown in Fig. 4(b) supports our highest-measured $Q$ factor of ~1800. This value is still much lower than our simulated value (~$8\times10^6$), likely due to light scattering by structural imperfections and/or optical absorption. To evaluate the variations of air hole positions and radii in the fabricated cavities, we fit the air hole edges in the SEM image of Fig. 2 (c) with circles [50]. By including the variations of the air hole positions (~5 nm standard deviation of $x$- and $y$-directions) and radii (~2.5 nm standard deviation) extracted from the SEM image into our FDTD simulations, we find the predicted $Q$ factor reduces to ~$8\times10^4$, which is still higher than the experimentally measured $Q$ factors. This suggests that other fabrication imperfections, such as the surface roughness of the slab, sidewall roughness, and tilt of the air holes likely play a significant role [53]. Note that the measured fluctuations of air hole positions and radii could be much smaller since extracted values contain the influence of charging in SEM measurements. We also noticed that there are tiny etch pits and undetermined contamination on the surface of the thin membrane, as seen in Figs. 3 (a)-(c). We suspect that this could induce additional undesired scattering loss and optical absorption. We believe that further optimization of the fabrication process and the use of surface cleaning techniques [54] can improve experimental $Q$ factors in the future.

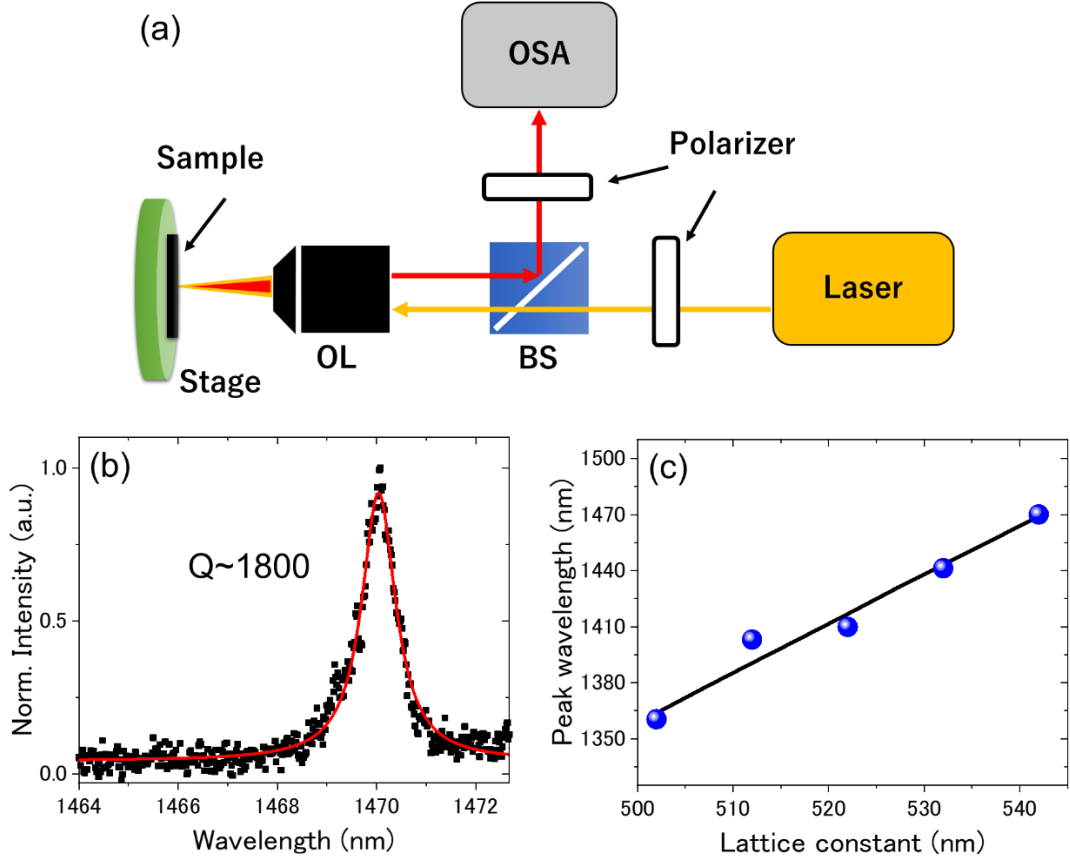

**Fig. 4.** (a) Illustration of the optical measurement setup. OL: objective lens; BS: beam splitter; OSA: optical spectrum analyzer. (b) Measured reflectance spectrum of the fabricated cavity with $a$ = 542 nm (black squares), overlaid with a fit using a Lorentzian function (red curve). (c) Extracted cavity resonances by fitting (blue dots) as a function of lattice constant $a$. The black solid line shows the simulated resonance wavelengths by the FDTD calculations.

 

In summary, we demonstrated telecommunication-wavelength 2D PhC cavities in a thin single-crystal diamond membrane. The 2D diamond PhC cavities can support a high theoretical $Q$ factor of ~$8 \times 10^6$ and a relatively small $V$ of ~2 $(\lambda/n)^3$. We developed a fabrication process using only standard EB lithography and top-down RIE for fabricating the cavities on the suspended thin diamond membrane. By resonant scattering spectroscopy, we observed cavity modes at wavelengths between 1360 and 1470 nm (telecommunication O- and S-band) with $Q$ factors up to 1800. Our fabrication technique using thin diamond membranes could enable large-scale fabrication of nanophotonic devices for various diamond-based photonic applications, involving nonlinear optics, optomechanics and sensing. In particular, high-$Q$ diamond 2D PhC cavities at telecommunication wavelengths could lead to low-threshold Raman lasers [31,32], optical parametric oscillators [55], or optical frequency converters [56].

**Acknowledgments**

We would like to thank Professor Iwamoto for his technical support. This work was supported by AFOSR (Grant Nos. FA9550- 19-1-0376 and FA9550-20-1-0105), ARO MURI (Grant No. W911NF1810432), NSF RAISE TAQS (Grant No. ECCS-1838976), NSF STC (Grant No. DMR-1231319), NSF ERC (Grant No. EEC- 1941583), DOE (Grant No. DE-SC0020376), ONR (Grant No. N00014-20-1-2425), Air Force (Grant No. FA8750-20-P-1716), and Australian Research Council Linkage Grants (Grant Nos. LP160101515 and LP190100528). K.K. acknowledges financial support from JSPS Overseas Research Fellowships (Project No. 202160592). D.R. acknowledges support from the NSF GRFP and Ford Foundation fellowships. A.S. acknowledges financial support from ARC DECRA Fellowship (No. DE190100336). N.S. acknowledges support of the Natural Sciences and Engineering Research Council of Canada (NSERC), and AQT Intelligent Quantum Networks and Technologies (INQNET) research program. The cavity simulation was performed at the University of Tokyo. Device fabrication was performed at the Center for Nanoscale Systems (CNS), a member of the National Nanotechnology Coordinated Infrastructure Network (NNCI), which is supported by the National Science Foundation under NSF Award No. 1541959. CNS is part of Harvard University.

**References**

[1] T. Yoshie, A. Scherer, J. Hendrickson, G. Khitrova, H.M. Gibbs, G. Rupper, C. Ell, O.B. Shchekin, and D.G. Deppe, Nature **432**, 200 (2004).

[2] K. Takeda, T. Sato, A. Shinya, K. Nozaki, W. Kobayashi, H. Taniyama, M. Notomi, K. Hasebe, T. Kakitsuka, and S. Matsuo, Nat. Photonics **7**, 569 (2013).

[3] M. Notomi, A. Shinya, S. Mitsugi, G. Kira, E. Kuramochi, and T. Tanabe, Opt. Express **13**, 2678 (2005).

[4] A.H. Safavi-Naeini, J.T. Hill, S. Meenehan, J. Chan, S. Gröblacher, and O. Painter, Phys. Rev. Lett. **112**, 153603 (2014).

[5] H. Kim, R. Bose, T.C. Shen, G.S. Solomon, and E. Waks, Nat. Photonics **7**, 373 (2013).

[6] E. Chow, A. Grot, L.W. Mirkarimi, M. Sigalas, and G. Girolami, Opt. Lett. **29**, 1093 (2004).

[7] J.P. Covey, A. Sipahigil, S. Szoke, N. Sinclair, M. Endres, and O. Painter, Phys. Rev. Appl. **11**, 034044 (2019).

[8] P.B. Deotare, M.W. McCutcheon, I.W. Frank, M. Khan, and M. Lončar, Appl. Phys. Lett. **94**, 121106 (2009).

[9] T. Asano, Y. Ochi, Y. Takahashi, K. Kishimoto, and S. Noda, Opt. Express **25**, 1769 (2017).

[10] L.-D. Haret, T. Tanabe, E. Kuramochi, and M. Notomi, Opt. Express **17**, 21108 (2009).

[11] M.J. Burek, Y. Chu, M.S.Z. Liddy, P. Patel, J. Rochman, S. Meesala, W. Hong, Q. Quan, M.D. Lukin, and M. Lončar, Nat. Commun. **5**, 5718 (2014).

[12] B.-S. Song, T. Asano, S. Jeon, H. Kim, C. Chen, D.D. Kang, and S. Noda, Optica **6**, 991 (2019).

[13] S. Sergent, M. Arita, S. Kako, K. Tanabe, S. Iwamoto, and Y. Arakawa, Phys. Status Solidi **10**, 1517 (2013).

[14] M. Khan, T. Babinec, M.W. McCutcheon, P. Deotare, and M. Lončar, Opt. Lett. **36**, 421 (2011).

[15] J. Riedrich-Möller, L. Kipfstuhl, C. Hepp, E. Neu, C. Pauly, F. Mücklich, A. Baur, M. Wandt, S. Wolff, M. Fischer, S. Gsell, M. Schreck, and C. Becher, Nat. Nanotechnol. **7**, 69 (2012).

[16] A. Faraon, C. Santori, Z. Huang, V.M. Acosta, and R.G. Beausoleil, Phys. Rev. Lett. **109**, 033604 (2012).

[17] B.J.M. Hausmann, B.J. Shields, Q. Quan, Y. Chu, N.P. de Leon, R. Evans, M.J. Burek, A.S. Zibrov, M. Markham, D.J. Twitchen, H. Park, M.D. Lukin, and M. Lončar, Nano Lett. **13**, 5791 (2013).

[18] L. Li, T. Schröder, E.H. Chen, M. Walsh, I. Bayn, J. Goldstein, O. Gaathon, M.E. Trusheim, M. Lu, J. Mower, M. Cotlet, M.L. Markham, D.J. Twitchen, and D. Englund, Nat. Commun. **6**, 6173 (2015).

[19] J. Riedrich-Möller, C. Arend, C. Pauly, F. Mücklich, M. Fischer, S. Gsell, M. Schreck, and C. Becher, Nano Lett. **14**, 5281 (2014).

[20] J.L. Zhang, S. Sun, M.J. Burek, C. Dory, Y. Tzeng, K.A. Fischer, Y. Kelaita, K.G. Lagoudakis, M. Radulaski, Z. Shen, N.A. Melosh, S. Chu, M. Lončar, and J. Vučković, Nano Lett. **18**, 1360 (2018).

[21] A. Sipahigil, R.E. Evans, D.D. Sukachev, M.J. Burek, J. Borregaard, M.K. Bhaskar, C.T. Nguyen, J.L. Pacheco, H.A. Atikian, C. Meuwly, R.M. Camacho, F. Jelezko, E. Bielejec, H. Park, M. Lončar, and M.D. Lukin, Science **354**, 847 (2016).

[22] R.E. Evans, M.K. Bhaskar, D.D. Sukachev, C.T. Nguyen, A. Sipahigil, M.J. Burek, B. Machielse, G.H. Zhang, A.S. Zibrov, E. Bielejec, H. Park, M. Lončar, and M.D. Lukin, Science **362**, 662 (2018).

[23] K. Kuruma, B. Pingault, C. Chia, D. Renaud, P. Hoffmann, S. Iwamoto, C. Ronning, and M. Lončar, Appl. Phys. Lett. **118**, 230601 (2021).

[24] A.E. Rugar, S. Aghaeimeibodi, D. Riedel, C. Dory, H. Lu, P.J. McQuade, Z. Shen, N.A. Melosh, and J. Vučković, Phys. Rev. X **11**, 031021 (2021).

[25] M.J. Burek, J.D. Cohen, S.M. Meenehan, N. El-Sawah, C. Chia, T. Ruelle, S. Meesala, J. Rochman, H.A. Atikian, M. Markham, D.J. Twitchen, M.D. Lukin, O. Painter, and M. Lončar, Optica **3**, 1404 (2016).

[26] J. V. Cady, O. Michel, K.W. Lee, R.N. Patel, C.J. Sarabalis, A.H. Safavi-Naeini, and A.C. Bleszynski Jayich, Quantum Sci. Technol. **4**, 024009 (2019).

[27] S.M. Meenehan, J.D. Cohen, S. Gröblacher, J.T. Hill, A.H. Safavi-Naeini, M. Aspelmeyer, and O. Painter, Phys. Rev. A **90**, 011803 (2014).

[28] H. Ren, M.H. Matheny, G.S. MacCabe, J. Luo, H. Pfeifer, M. Mirhosseini, and O. Painter, Nat. Commun. **11**, 3373 (2020).

[29] G.S. MacCabe, H. Ren, J. Luo, J.D. Cohen, H. Zhou, A. Sipahigil, M. Mirhosseini, and O. Painter, Science **370**, 840 (2020).

[30] S. Maity, L. Shao, S. Bogdanović, S. Meesala, Y.-I. Sohn, N. Sinclair, B. Pingault, M. Chalupnik, C. Chia, L. Zheng, K. Lai, and M. Lončar, Nat. Commun. **11**, 193 (2020).

[31] Y. Takahashi, Y. Inui, M. Chihara, T. Asano, R. Terawaki, and S. Noda, Nature **498**, 470 (2013).

[32] P. Latawiec, V. Venkataraman, M.J. Burek, B.J.M. Hausmann, I. Bulu, and M. Lončar, Optica **2**, 924 (2015).

[33] T. Jung, L. Kreiner, C. Pauly, F. Mücklich, A.M. Edmonds, M. Markham, and C. Becher, Phys. Status Solidi **213**, 3254 (2016).

[34] J.C. Lee, A.P. Magyar, D.O. Bracher, I. Aharonovich, and E.L. Hu, Diam. Relat. Mater. **33**, 45 (2013).

[35] B.A. Fairchild, P. Olivero, S. Rubanov, A.D. Greentree, F. Waldermann, R.A. Taylor, I. Walmsley, J.M. Smith, S. Huntington, B.C. Gibson, D.N. Jamieson, and S. Prawer, Adv. Mater. **20**, 4793 (2008).

[36] O. Gaathon, J.S. Hodges, E.H. Chen, L. Li, S. Bakhru, H. Bakhru, D. Englund, and R.M. Osgood, Opt. Mater. (Amst). **35**, 361 (2013).

[37] A.P. Magyar, J.C. Lee, A.M. Limarga, I. Aharonovich, F. Rol, D.R. Clarke, M. Huang, and E.L. Hu, Appl. Phys. Lett. **99**, 081913 (2011).

[38] B. Khanaliloo, M. Mitchell, A.C. Hryciw, and P.E. Barclay, Nano Lett. **15**, 5131 (2015).

[39] S. Mouradian, N.H. Wan, T. Schröder, and D. Englund, Appl. Phys. Lett. **111**, 021103 (2017).

[40] N.H. Wan, S. Mouradian, and D. Englund, Appl. Phys. Lett. **112**, 141102 (2018).

[41] A.H. Piracha, K. Ganesan, D.W.M. Lau, A. Stacey, L.P. McGuinness, S. Tomljenovic-Hanic, and S. Prawer, Nanoscale **8**, 6860 (2016).

[42] E. Kuramochi, M. Notomi, S. Mitsugi, A. Shinya, T. Tanabe, and T. Watanabe, Appl. Phys. Lett. **88**, 041112 (2006).

[43] Z. Han, X. Checoury, L.-D. Haret, and P. Boucaud, Opt. Lett. **36**, 1749 (2011).

[44] M. Minkov, U.P. Dharanipathy, R. Houdré, and V. Savona, Opt. Express **21**, 28233 (2013).

[45] S. Tomljenovic-Hanic, A.D. Greentree, C.M. de Sterke, and S. Prawer, Opt. Express **17**, 6465 (2009).

[46] A.H. Piracha, P. Rath, K. Ganesan, S. Kühn, W.H.P. Pernice, and S. Prawer, Nano Lett. **16**, 3341 (2016).

[47] I. Aharonovich, J.C. Lee, A.P. Magyar, B.B. Buckley, C.G. Yale, D.D. Awschalom, and E.L. Hu, Adv. Mater. **24**, OP54 (2012).

[48] Y. Tao and C. Degen, Adv. Mater. **25**, 3962 (2013).

[49] M. Salz, Y. Herrmann, A. Nadarajah, A. Stahl, M. Hettrich, A. Stacey, S. Prawer, D. Hunger, and F. Schmidt-Kaler, Appl. Phys. B **126**, 131 (2020).

[50] K. Kuruma, Y. Ota, M. Kakuda, S. Iwamoto, and Y. Arakawa, APL Photonics **5**, 046106 (2020).

[51] A.R.A. Chalcraft, S. Lam, D. O'Brien, T.F. Krauss, M. Sahin, D. Szymanski, D. Sanvitto, R. Oulton, M.S. Skolnick, A.M. Fox, D.M. Whittaker, H.-Y. Liu, and M. Hopkinson, Appl. Phys. Lett. **90**, 241117 (2007).

[52] T. Tajiri, Y. Sakai, K. Kuruma, S.M. Ji, H. Kiyama, A. Oiwa, J. Ritzmann, A. Ludwig, A.D. Wieck, Y. Ota, Y. Arakawa, and S. Iwamoto, Jpn. J. Appl. Phys. **59**, SGGI05 (2020).

[53] T. Asano, B. Song, and S. Noda, Opt. Express **14**, 1996 (2006).

[54] O.M. Suprun, G.D. Il'nyts'ka, V.M. Tkach, and S.O. Ivakhnenko, J. Superhard Mater. **37**, 211 (2015).

[55] B.J.M. Hausmann, I. Bulu, V. Venkataraman, P. Deotare, and M. Loncar, Nat. Photonics **8**, 369 (2014).

[56] Q. Shen, A. Shams-Ansari, A.M. Boyce, N.C. Wilson, T. Cai, M. Loncar, and M.H. Mikkelsen, Nanophotonics **10**, 589 (2020).